\documentstyle[12pt]{article}
\begin{document}
\thispagestyle{empty}
\begin{center}
\LARGE \tt \bf {On the stability of cosmological metrics in spacetimes with spin-torsion fluctuations}
\end{center}
\vspace{2.5cm}
\begin{center} {\large L.C. Garcia de Andrade\footnote{Departamento de Fisica Teorica-Instituto de Fisica-Universidade do Estado do Rio de Janeiro-UERJ-Rua Sao Francisco Xavier-524-Rio de Janeiro, Brasil.}}
\end{center}
\vspace{2cm}
\begin{abstract}
Two metric perturbations in Einstein-Cartan cosmology are examined.The first case is the scalar
mode perturbation of de Sitter metric in Einstein-Cartan cosmology.In this case small perturbations of de Sitter metric shows that this metric is unstable for a universe with spin-torsion density and dilaton fields.In the second case we show that the Friedmann metric in the same space is also unstable under the small perturbations.
\end{abstract}
\newpage
The investigation of stability of cosmological metrics have been proved to be important in the understanding of formation structure like galaxy and star formation \cite{1,2,3}.Interesting example as the stability of the Godel model rotating universe along the radial direction and the instability of the same metric along the rotation axis.This result was proved by Silk \cite{4} in the realm of general relativistic cosmologies.From the point of view of torsion theories some two interesting examples have been provided.The first was given by Nurgaliev and Ponomariev \cite{5} who showed that the Friedmann metric could be stable under small homogeneous and isotropic perturbations.On the other hand more recently Maroto and Shapiro \cite{6} have shown that de Sitter metric in D-dimensional higher order gravity with dilatons and torsion could be always unstable under perturbations.In this letter to show that this result is still valid by showing that the Friedmann and de Sitter metric are both unstable in the case of EC dilatonic cosmology.It is seems interesting to point it out however that in Maroto and Shapiro case torsion is not produced by spin but it is a string torsion although it does not propagate,or is represented by an algebraic equation exactly as in the case of EC cosmology considered here and in Ponomariev and Nurgaliev paper.The method of perturbation considered by Maroto and Shapiro takes into account the action of the inflationary cosmology in D-dimensional gravity and the Lagrangean method of investigating the stability.Let us now consider the perturbed Friedmann metric  
\begin{equation}
ds^{2}=dt^{2}-a^{2}(t)({\delta}_{ik}+h_{ik})dx^{i}dx^{k} 
\label{1}
\end{equation}
where ${i,k=1,2,3}$ and ${\delta}_{ik}$ is the Kronecker delta.In the case the background metric is de Sitter $a(t)=e^{H_{0}t}$ where $H_{0}$ is the Hubble constant and $t$ is the cosmic time.Besides $H_{0}=\frac{\dot{a}}{a}$.The basic equation of the theory is equation (\ref{16}) bellow.The Ricci tensor perturbed is 
\begin{equation}
R'_{00}=R_{00}+{\delta}R_{00}
\label{2}
\end{equation}
where
\begin{equation}
{\delta}R_{00}=\frac{1}{a^{2}}[{\ddot{h}}_{kk}-2\frac{\dot{a}}{a}{\dot{h}}_{kk}+2(\frac{{\dot{a}}^{2}}{a^{2}}-\frac{\ddot{a}}{a})h_{kk}]
\label{3}
\end{equation}
where $h_{kk}=a^{2}h$ and the energy-momentum tensor is given by
\begin{equation}
{T'}_{00}=T_{00}+{\delta}{\rho}_{eff}
\label{4}
\end{equation}
where ${\rho}_{eff}={\rho}-2{\pi}G{\sigma}^{2}$ where ${\sigma}^{2}$ is the spin-torsion density in order that we can consider unpolarized spinning matter in any phase of the Universe.Energy-momentum tensor trace is given by  
\begin{equation}
T'=T+{\delta}T=T+{\delta}{\rho}_{eff}
\label{5}
\end{equation}
Here we consider a dust like universe with spin-torsion density and massless dilatons given by the scalar field ${\phi}$ where  ${\rho}=\frac{1}{2}{\dot{\phi}}^{2}$ and since the dilatons are throughout this work massless the dilaton potential $V({\phi})$ vanishes leaving only the dilaton kinetic energy.The background de Sitter metric obeys the following Einstein-Cartan equations
\begin{equation}
{H_{0}}^{2}=\frac{8{\pi}G}{3}({\rho}_{0}-2{\pi}G{{\sigma}_{0}}^{2})
\label{6}
\end{equation}
and
\begin{equation}
H^{2}=\frac{8{\pi}G}{3}({\rho}-2{\pi}G{\sigma}^{2})
\label{7}
\end{equation}
while the dilaton field equation is given by
\begin{equation}
{\ddot{\phi}}+H_{0}{\dot{\phi}}=0
\label{8}
\end{equation}
The perturbed Einstein-Cartan field equations for the dilaton system is 
\begin{equation}
{\delta}R_{ab}=8{\pi}G({\delta}T_{ab}-\frac{1}{2}{\delta}T)
\label{9}
\end{equation}
where ${a,b=0,1,2,3}$.To compute the RHS of the energy-momentum tensor we have first to obtain  the solution for the massless dilaton which is 
\begin{equation}
{\phi}=e^{-H_{0}t}
\label{10}
\end{equation}
which allows us to obtain the perturbation on the dilaton field ${\dot{\phi}}{\delta}{\dot{\phi}}$ and substitute this result into
\begin{equation}
{\delta}T_{00}-\frac{1}{2}{\delta}T={\delta}{\dot{\phi}}^{2}-2{\pi}G{\delta}{\sigma}^{2}
\label{11}
\end{equation}
where we have used the fact that ${\delta}\frac{d}{dt}=\frac{d}{dt}{\delta}$ and where to obtain this equation we have made use of the dilaton background solution \begin{equation}
{\dot{\phi}}^{2}=\frac{1}{{\pi}G}{\sigma}^{2}
\label{12}
\end{equation}
Substitution of this last expression on the perturbed Einstein-Cartan equation yields the
evolution equation of the dynamics of the scalar mode perturbation $h(t)$ 
\begin{equation}
\ddot{h}=\frac{1}{4}(1-2({\pi}G)^{2})e^{-2H_{0}t}
\label{13}
\end{equation}
where we have approximate the equations for values of time very far away from the observer.The solution of this last equation yields
\begin{equation}
h(t)=\frac{(1-2({\pi}G)^{2})e^{-2H_{0}t}}{{H_{0}}^{2}}+At
\label{14}
\end{equation}
where A is an integration constant.Since the first term on the RHS of the scalar mode of perturbation equation vanishes as the time evolves but the second term grows linearly faster than the first term decays one may conclude that de Sitter metric is unstable under small perturbation.To investigate and show that this result is valid also in the Friedmann case we need to make some assumptions to simplify the matters.The first is that we have a two fluid universe where the inner fluid represents a general relativistic Friedmann model which obeys the equation 
\begin{equation}
H^{2}_{0}=\frac{8{\pi}G}{3}{\rho}
\label{15}
\end{equation}
While the outer metric would be given by a Friedmann model in EC cosmology which equation would be
\begin{equation}
H_{1}^{2}+\frac{1}{a^{2}}=\frac{8{\pi}G}{3}({\rho}-2{\pi}G{\sigma}^{2})
\label{16}
\end{equation}
These formulas allow us to obtain the density perturbation $\frac{{\delta}{\rho}}{\rho}$ which we shall need later in the letter.Thus in the case of superhorizon one obtains
\begin{equation}
\frac{{\delta}{\rho}}{\rho}=-(\frac{3}{8{\pi}Ga^{2}{\rho}}+\frac{{2{\pi}G}{\delta}{\sigma}^{2}}{\rho})
\label{17}
\end{equation}
It is easy to check that the RHS of the equation (\ref{17}) is compatible with the linear perturbation condition $\frac{{\delta}{\rho}}{\rho}<<1$ by writing the RHS of this same equation in terms of the redshift z.This is simple if one notices that $a=(1+z)^{-1}$ and that the density ${\rho}=Ma^{-3}$ where M is the mass of the stellar object.Therefore formula (\ref{17}) may be valid even for low redshifts.Substitution of these expressions into equation (\ref{17}) one obtains
\begin{equation}
\frac{{\delta}{\rho}}{\rho}=-(\frac{3}{8{\pi}G}\frac{(1+z)^{-1}}{M}+{2{\pi}G}{\delta}{\sigma}^{2}(1+z)^{-3})
\label{18}
\end{equation}
Thus for high redshifts objects $z=10^{3}$ we note that the RHS of expression (\ref{17}) is compatible with the LHS since in this case $z>>1$.Since in the dilatonic case ${\rho}={\dot{\phi}}^{2}$ one has that $\frac{{\delta}{\rho}}{\rho}=\frac{{\delta}{\dot{\phi}}^{2}}{{\phi}^{2}}$.By substitution of this result into equation (\ref{17}) one obtains 
\begin{equation}
({\delta}{\dot{\phi}}^{2}-2{\pi}G{\sigma}^{2})=\frac{3}{8{\pi}Ga^{2}}
\label{19}
\end{equation}
In this equation also the RHS is small as the RHS since we are considering the universe at stages where the cosmic scale a is too big or almost present universe,besides the value of the spin-torsion density ${\sigma}^{2}$ is of the order $10^{-30} gcm^{-13}$ at the present universe (see reference {\ref{10}) for details.Now let us notice that in the Friedmann case the perturbation equation above reads
\begin{equation} 
\ddot{h}+2\frac{\dot{a}}{a}\dot{h}+\frac{\ddot{a}}{a}h=({\delta}{\dot{\phi}}^{2}-2{\pi}G{\sigma}^{2})
\label{20}
\end{equation}
since the interior metric is GR metric one may use the general relativistic results $a=t^{\frac{2}{3}}$ into expression (\ref{19}) and subsequent substitution into (\ref{20}) yields
\begin{equation}
\ddot{h}+\frac{4}{3t}\dot{h}-\frac{1}{3t^{2}}h=\frac{3}{8{\pi}Gt^{\frac{4}{3}}}
\label{21}
\end{equation}
A simple solution of this equation can be found by assuming a solution of the type $h=At^{p}$ one may solve equation (\ref{20}) to yield
\begin{equation}
h(t)=\frac{9}{8{\pi}G}t^{\frac{2}{3}}+Bt^{\frac{-1\pm\sqrt{13}}{6}}
\label{22}
\end{equation}
where B is another integration constant.This expression shows finally our result that the Friedmann metric in EC cosmology is unstable under the assumptions we have made,since both terms on the RHS of expression (\ref{22}) are growing scalar modes in cosmic time $t$.Nurgaliev has some years ago made a detailed study of the structural stability of the Friedmann metric in EC cosmology \cite{7}.More recently density and metric perturbations in Brans-Dicke gravity has been considered by Fabris et al \cite{8}.To resume we have shown that de Sitter and Friedmann metrics can be unstable in EC dilatonic cosmology with spin-torsion density fluctuations \cite{9} as long as certain assumptions are made.A more detailed and general account of the topic discussed here may be done elsewhere.
\section*{Acknowledgments}
\paragraph*{}
We are very much indebt to Professors Ilya Shapiro and Patricio Letelier   for enlightening discussions on the subject of this paper.Thanks are also due to an unknown referee for very useful comments and suggestions on the first draft of this letter.Financial supports from CNPq. (Brazilian Government Agency), Universidade do Estado do Rio de Janeiro (UERJ) and FAPESP are gratefully acknowledged. 

\end{document}